\documentclass[onefignum,onetabnum]{siamart190516}

\usepackage{graphicx}
\usepackage{graphicx}
\usepackage{amsmath}
\usepackage{amssymb}
\usepackage{xcolor}
\usepackage[square,sort,comma,numbers]{natbib}
\usepackage{hyperref}
\usepackage{ulem}
\usepackage{bbm}
\ifpdf
\DeclareGraphicsExtensions{.eps,.pdf,.png,.jpg}
\else
\DeclareGraphicsExtensions{.eps}
\fi

\newsiamremark{remark}{Remark}
\newsiamremark{hypothesis}{Hypothesis}
\crefname{hypothesis}{Hypothesis}{Hypotheses}
\newsiamthm{claim}{Claim}

\headers{Stochastic Heteroclinic Cycling}{N. W. Barendregt and P. J. Thomas}

\title{Heteroclinic Cycling and Extinction in May-Leonard Models with Demographic Stochasticity
	\thanks{\funding{This work was made possible in part by grants from the National Science Foundation (DMS-2052109 and DEB-1654989). 
			This research has been supported in part by the Mathematical Biosciences Institute and the National Science Foundation under grant DMS-1440386.}}}

\author{Nicholas W. Barendregt
	\thanks{Department of Applied Mathematics, University of Colorado Boulder, Boulder, CO (\email{nicholas.barendregt@colorado.edu}).}
	\and Peter J. Thomas
	\thanks{Department of Mathematics, Applied Mathematics, and Statistics, Department of Biology, Department of Cognitive Science, Department of Data and Computer Science, Department of Electrical, Computer and Systems Engineering, Case Western Reserve University, Cleveland, OH (\email{pjthomas@case.edu}).}}

\usepackage{amsopn}

\ifpdf
\hypersetup{
	pdftitle={Stochastic Heteroclinic Cycling},
	pdfauthor={N. W. Barendregt and P. J. Thomas}
}
\fi

\begin{document}
	
	\maketitle
	
	% REQUIRED
	\begin{abstract}
		May and Leonard (SIAM J.~Appl.~Math 1975) introduced a three-species Lotka-Volterra type population model that exhibits heteroclinic cycling.  
		Rather than producing a periodic limit cycle, the trajectory takes longer and longer to complete each ``cycle", passing closer and closer to unstable fixed points in which one population dominates and the others approach zero. 
		Aperiodic heteroclinic dynamics have subsequently been studied in ecological systems (side-blotched lizards; colicinogenic E. coli), in the immune system, in neural information processing models (``winnerless competition"), and in models of neural central pattern generators. 
		Yet as May and Leonard observed ``Biologically, the behavior (produced by the model) is nonsense. Once it is conceded that the variables represent animals, and therefore cannot fall below unity, it is clear that the system will, after a few cycles, converge on some single population, extinguishing the other two."
		Here, we explore different ways of introducing discrete stochastic dynamics based on May and Leonard's ODE model, with application to ecological population dynamics, and to a neuromotor central pattern generator system. 
		We study examples of several quantitatively distinct asymptotic behaviors, including total extinction of all species, extinction to a single species, and persistent cyclic dominance with finite mean cycle length.
	\end{abstract}
	
	% REQUIRED
	\begin{keywords}
		Stochastic Modeling, Heteroclinic Cycling, Theoretical Ecology, Computational Neuroscience
	\end{keywords}
	
	% REQUIRED
	\begin{AMS}
		92B05, 37C29, 60J27, 60J22
	\end{AMS}
	
	\section{Introduction}
	Following Lotka \cite{lotka1925} and Volterra \cite{volterra1926}, May and Leonard \cite{May1975} introduced a model  generalizing Lotka-Volterra dynamics for a system of three species:
	\begin{equation}
		\begin{aligned}
			\frac{dn_1}{dt}={}&n_1\left(1-n_1-\alpha n_2-\beta n_3\right),\\
			\frac{dn_2}{dt}={}&n_2\left(1-\beta n_1-n_2-\alpha n_3\right),\\
			\frac{dn_3}{dt}={}&n_3\left(1-\alpha n_1-\beta n_2-n_3\right).
		\end{aligned}
		\label{eq:HC_Deterministic}
	\end{equation}
	In \Cref{eq:HC_Deterministic} $n_i$ represents the population of species $i$, and the constants $\alpha\ge0$ and $\beta\ge0$ represent the strengths of competitive interactions.
	The model exhibits different types of coexistence for different choices of $\alpha$ and $\beta$.
	When $\alpha+\beta=2$, the system converges to a periodic orbit contained in the plane $n_1+n_2+n_3=1$. 
	This solution can be interpreted as the direct extension of Lotka-Volterra to three species, where each species' population oscillates with finite period.
	However, when $\alpha+\beta>2$ and either $\alpha < 1$ or $\beta < 1$, the system undergoes heteroclinic cycling, with the duration of each cycle increasing as time progresses.
	In this regime, each species' population becomes closer to zero with each cycle, and spends a longer fraction of each cycle in this near-extinction state.
	
	Heteroclinic cycling models such as \Cref{eq:HC_Deterministic} and their variants
	have frequently served as models for ``rock-paper-scissors''-type population dynamics in which populations take turns as the dominant species before being pushed out in favor of a more competitive population. 
	Sinervo and  Lively \cite{Sinervo1996} found that a species of side-blotched lizards exhibits rock-paper-scissors competition: orange aggressive lizards beat out less-aggressive blue lizards for mates, yellow ``sneaker'' lizards invade the larger orange lizard territory to steal mates, and blue lizards beat out the sneakers for mates.   
	Kerr et. al. \cite{Kerr2002} observed a similar behavior in colicinogenic \textit{E. coli}: a toxin-producing strain kills a susceptible population, a toxin-resistant population grows faster than the toxin-producing population, and the susceptible population grows faster than the resistant population.
	In computational neuroscience, heteroclinic cycling has been proposed as an alternative to classic ``winner-take-all'' models for neural networks.
	Rabinovich et. al. \cite{Rabinovich2001,rabinovich2008} suggested that the activity of olfactory neurons when encoding stimuli can be projected onto a heteroclinic cycle and called the behavior ``winnerless competition.''
	Varona et. al. \cite{varona2002} theorized that high-dimensional heteroclinic systems leading to chaotic dynamics might underlie the apparently random search behavior during hunting in the mollusc \textit{Clione}.
	Shaw et. al. \cite{shaw15} and Lyttle et. al. \cite{Lyttle2017} constructed a model capable of transitioning between limit-cycling and heteroclinic-cycling behaviors to represent a neuromotor central pattern generator (CPG) in  \textit{Aplysia californica} (see also \cite{park2018}). 
	While more  detailed models for the \textit{Aplysia} feeding system have since been developed \cite{webster2020}, the simplicity of the three-component SLG (Shaw-Lyttle-Gill) model makes it an attractive target for analysis.  
	
	Despite their popularity, heteroclinic cycling models of biological populations, when formulated as systems of ordinary differential equations, suffer a fundamental flaw.
	Indeed, in their original paper, May and Leonard noted a significant drawback of their model's ability to describe population dynamics.
	They observed that, while heteroclinic cycling continues indefinitely, real biological populations ``cannot fall below unity, [and] it is clear that the system will, after a few cycles, converge on some single population, extinguishing the other two'' \cite{May1975}.
	This discrepancy arises from demographic stochasticity, or copy number noise, that is inherent in systems where populations take on discrete integer values.
	
	In light of May and Leonard's observation, one might expect that a stochastic system undergoing heteroclinic cycling would necessarily exhibit population extinctions.
	However, as is well known, the mapping from a given ODE model to a stochastic model having matching mean-field dynamics is not unique.   
	For example, Allen \cite{allen2010} noted that for a logistic birth-death process, there are an infinite number of per capita birth and death rates that yield the same mean-field logistic growth. 
	Xue and Goldenfeld \cite{xue2017} found that modeling plankton ecosystems using stochastic versions of the ``kill-the-winner'' model resulted in extinction events, while the mean-field model had stable coexistence of all species. 
	And  Strang et. al. \cite{Strang2019} explored the paradox that stochastic models with the Allee effect, which reduces per-capita growth rate for small population size, can have longer persistence than models without the effect.
	The ambiguity intrinsic to stochastic extensions of ODE systems is not confined to ecological models.  
	A series of papers have debated the most appropriate way to extend the deterministic Hodgkin-Huxley equations to incorporate the effects of random gating of ion channels in neural dynamics \cite{fox1994,goldwyn2011,goldwyn2011a,orio2012,Anderson2015b,pu2020,pu2021}
	At the level of large-scale neural circuits, several distinct stochastic generalizations have been proposed that coincide with the classical deterministic Wilson-Cowan neural field equations in the mean-field limit \cite{bressloff2010,benayoun2010,faugeras2015,cowan2016,de2021}.
	
	As these examples suggest, there could be more than one stochastic model consistent with \Cref{eq:HC_Deterministic} in the mean-field limit, but exhibiting distinct long-term behaviors for finite system size.
	In this paper, we investigate three different stochastic implementations of heteroclinic cycling, each resulting in distinct long-term behavior.
	First, we consider two alternative stochastic models, each based on a birth-death formalism consistent with \Cref{eq:HC_Deterministic}. 
	By formulating the discrete master equation \cite{Gardiner09} and leveraging complex-balanced equilibrium results from chemical kinetics \cite{Anderson2015a}, we prove that each alternative results in a qualitatively different stationary distribution.
	We confirm these findings numerically.
	We then propose a modified May-Leonard system inspired by a neuromotor CPG model from Lyttle et. al. \cite{Lyttle2017}.
	Using the same birth-death formalism, we construct a stochastic implementation of this new model that not only avoids extinction events, but also maintains a finite mean cycle length.
	We numerically investigate how the mean cycle length depends on model parameters and examine its asymptotic behaviors in the both the large and small system size limits.
	Taken together, these results illustrate the rich variety of behaviors that may be obtained from different stochastic generalizations of May and Leonard's original deterministic heteroclinic cycling model.
	
	\section{Mean-Field Formulations of Heteroclinic Cycling}
	For a general system of $m$ species following deterministic Lotka-Volterra interactions,  species $i$ has the governing equation
	\begin{equation}
		\frac{dn_i}{dt}=r_in_i\left(1-\sum_{j=1}^{m}k_{ij}n_j\right)+f_i(t).
		\label{eq:Generalized_Lotka-Volterra}
	\end{equation}
	In \Cref{eq:Generalized_Lotka-Volterra}, $n_i$ is the population size of species $i\in\{1,\dots,m\}$, $r_i$ is the intrinsic growth rate of species $i$, $k_{ij}$ represents the strength the competitive effect of species $j$ on species $i$, and $f_i(t)$ is a nonhomogeneous forcing function that can represent immigration, harvesting, etc.~of species $i$.
	We will use \Cref{eq:Generalized_Lotka-Volterra} to construct three versions of May and Leonard's heteroclinic cycling model.
	For the duration of the paper we will restrict our attention to three interacting species ($m=3$), assume that each species has the same intrinsic growth rate $r_1=r_2=r_3=r$ and forcing function $f_1=f_2=f_3=f$, and enforce that competition rates have the same cyclic symmetry as the May-Leonard system, so that $k_{12}=k_{23}=k_{31}=k_{i,i+1}$, $k_{13}=k_{21}=k_{32}=k_{i,i+2}$ and $k_{11}=k_{22}=k_{33}=k_{ii}$, where indicial addition is taken cyclically.
	Note that by setting $m=3$, $r=1$, $k_{i,i+1}=\alpha$, $k_{i,i+2}=\beta$, $k_{ii}=1$, and $f=0$, we recover \Cref{eq:HC_Deterministic}.
	
	The first two models we consider will be direct analogues of  \Cref{eq:HC_Deterministic}.
	As is the case in May and Leonard's original system, both models will obey mass-action kinetics, with implications that we discuss below.
	We begin with a ``general variance'' or ``GV model.'' In this model, the intrinsic growth rate $r$ reflects the combined effects of a per capita birth rate $b>0$ and a per capita death rate $d>0$, chosen so that $r=b-d>0$.
	The terminology ``general variance" reflects the fact that the variance of the population growth over short times $\Delta t$ scales as $(b+d)\Delta t+o(\Delta t)$. 
	Thus for a given value of $r$, we can obtain arbitrarily large variance in the population growth process by increasing both $b$ and $d$.
	Following the language of van Kampen \cite{van92} and Gardiner \cite{Gardiner09}, 
	we introduce a system size parameter $\Omega$ (representing the single-species carrying capacity).
	We consider the $n_i$ of \Cref{eq:Generalized_Lotka-Volterra} as intensive variables and define $N_i=\Omega n_i$ as extensive variables for the number of individuals in the $i$-th species.
	The resulting mean-field equations for the GV model may be written as:
	\begin{equation}
		\begin{aligned}
			\frac{dN_1}{dt} ={}& N_1\left[(b-d)-\frac{N_1}{\Omega}-\frac{\alpha}{\Omega} N_2-\frac{\beta}{\Omega} N_3\right], \\
			\frac{dN_2}{dt} ={}& N_2\left[(b-d)-\frac{\beta}{\Omega} N_1-\frac{N_2}{\Omega}-\frac{\alpha}{\Omega} N_3\right], \\
			\frac{dN_3}{dt} ={}& N_3\left[(b-d)-\frac{\alpha}{\Omega} N_1-\frac{\beta}{\Omega} N_2-\frac{N_3}{\Omega}\right].
		\end{aligned}
		\label{eq:GV_Model_MF}
	\end{equation}
	For notational clarity, we write the birth and death rates separately; in the stochastic model each will parametrize a separate stochastic reaction term (see \cref{sec:GV_Model}).
	Note that when $N_2=N_3=0$, $N_1$ follows logistic growth with carrying capacity $\Omega$ and low-density growth rate $(b-d)$.
	
	The second model we consider may be seen as a special case of the GV model, given by setting the intrinsic growth rate $r=b$ and the per capita death rate $d=0$.
	While this restriction may seem nonphysical, it may be a good approximation of some biological systems. For example, some bacterial populations survive exposure to antibiotics by entering a ``persistent state'' for which the mortality rate is effectively zero (see \cite{gerdes2012,browning2021} for details). 
	As noted above, the variance of the population growth over short times is proportional to $b+d$. Therefore, for a fixed $r$, the assumption $d=0$ gives the \emph{minimum variance model}, which we call the ``minimal model.''
	Its mean-field equations are:
	\begin{equation}
		\begin{aligned}
			\frac{dN_1}{dt} ={}& N_1\left[r-\frac{N_1}{\Omega}-\frac{\alpha}{\Omega} N_2-\frac{\beta}{\Omega} N_3\right], \\
			\frac{dN_2}{dt} ={}& N_2\left[r-\frac{\beta}{\Omega} N_1-\frac{N_2}{\Omega}-\frac{\alpha}{\Omega} N_3\right], \\
			\frac{dN_3}{dt} ={}& N_3\left[r-\frac{\alpha}{\Omega} N_1-\frac{\beta}{\Omega} N_2-\frac{N_3}{\Omega}\right].
		\end{aligned}
		\label{eq:Minimal_Model_MF}
	\end{equation}
	In \Cref{eq:Minimal_Model_MF}, we replaced the individual birth and death rates from the GV model with the net growth rate $r$.
	Again note that if $b-d=r$ from \Cref{eq:GV_Model_MF}, the GV and minimal models are equivalent at the level of mean-field equations, and we recover \Cref{eq:HC_Deterministic} by taking $\Omega=b-d=r\equiv1$.
	However, in the stochastic implementation of the minimal model, eliminating the death process qualitatively changes the long-time asymptotic behavior (see \cref{sec:Minimal_Model}).
	
	As a third stochastic variation on the May-Leonard model, we explore the effect of the nonhomogeneous term $f\neq0$.  
	This variation is motivated by heteroclinic cycling models of neural CPGs.
	For example, Shaw et. al. \cite{shaw15} and Lyttle et. al. \cite{Lyttle2017} proposed a model for a CPG driving feeding movements in the marine mollusk \textit{Aplysia californica} that comprises three pools of motor neurons, coupled by reciprocal inhibition and driven by endogenous activation.
	Each neural pool has an activation variable, $a_i$, $i\in\{0,1,2\}$, ranging from $a_i=0$ (inactive) to $a_i=1$ (fully active), and satisfying May-Leonard type competitive dynamics.  
	To study the effects of demographic stochasticity, we interpret the $a_i$ as intensive variables representing the fraction of active neurons in $i$-th pool, analogous to the Wilson-Cowan equations \cite{Wilson1972,Wilson1973}.  
	We introduce a system size $\Omega$, corresponding to the number of cells in each pool, and write $A_i=\Omega a_i$ as extensive variables, representing the integer number of active cells. We thus obtain our third mean-field model, which we call the ``three-pool model:''
	\begin{equation}
		\begin{aligned}
			\frac{dA_0}{dt}={}&\frac{1}{\tau}\left[A_0\left(1-\frac{A_0}{\Omega}-\frac{\gamma}{\Omega} A_1\right)+\mu\left(\Omega-A_0\right)\right],\\
			\frac{dA_1}{dt}={}&\frac{1}{\tau}\left[A_1\left(1-\frac{A_1}{\Omega}-\frac{\gamma}{\Omega} A_2\right)+\mu\left(\Omega-A_1\right)\right],\\
			\frac{dA_2}{dt}={}&\frac{1}{\tau}\left[A_2\left(1-\frac{A_2}{\Omega}-\frac{\gamma}{\Omega} A_0\right)+\mu\left(\Omega-A_2\right)\right].
		\end{aligned}
		\label{eq:Three_Pool_Deterministic}
	\end{equation}
	Note that \Cref{eq:Three_Pool_Deterministic} can be obtained from \Cref{eq:Generalized_Lotka-Volterra} by taking $r=\frac{1-\mu}{\tau}$, $k_{ii}=\frac{1}{\tau\Omega}$, $k_{i,i+1}=\frac{\gamma}{\tau\Omega}$, $k_{i,i+2}=0$, and $f=\frac{\mu\Omega}{\tau}$.
	In \Cref{eq:Three_Pool_Deterministic}, $\tau$ is a time constant, $\gamma$ is the strength of inhibition, and $\mu$ governs the rate of endogenous activation.
	This activation parameter $\mu$ represents intrinsic sources of excitation, whether from ongoing network activity, slow endogenous excitatory currents, or neuromodulatory effects, that cause cells to activate spontaneously. 
	This endogenous activation provides an additional source of stochasticity in our model.
	In this model, the total number of cells in each neural pool is conserved, with transitions representing changes of activation state rather than ``births" or ``deaths".
	In contrast to the ecological models, \Cref{eq:GV_Model_MF} and \Cref{eq:Minimal_Model_MF}, where population sizes are unbounded, in the neural pool model the population state-space finite.
	The endogenous activation term $\mu\ll 1$ was introduced by Shaw et. al. \cite{shaw15} as a means of regulating the sensitivity of the neural activity, by steering trajectories away from the saddle points of the heteroclinic system.  
	Here we define the endogenous activation term somewhat differently from their original formulation, in order to enforce zero flux conditions on the boundaries of our space, which in turn allows us to construct a well-defined stochastic model (see \cref{sec:Three_Pool_Model}).
	As in the GV and minimal models, the three-pool model obeys mass-action kinetics; to see this, define $I_i=\Omega-A_i$ to be the number of inactive neurons in the $i$-th pool.
	We may then rewrite \Cref{eq:Three_Pool_Deterministic} as:
	\begin{equation*}
		\begin{aligned}
			\frac{dA_0}{dt}={}&\frac{1}{\tau}\left[\frac{1}{\Omega}A_0I_0-\frac{\gamma}{\Omega}A_0A_1+\mu I_0\right],\\
			\frac{dA_1}{dt}={}&\frac{1}{\tau}\left[\frac{1}{\Omega}A_1I_1-\frac{\gamma}{\Omega}A_1A_2+\mu I_1\right],\\
			\frac{dA_2}{dt}={}&\frac{1}{\tau}\left[\frac{1}{\Omega}A_2I_2-\frac{\gamma}{\Omega}A_2A_0+\mu I_2\right].
		\end{aligned}
	\end{equation*}
	
	\begin{figure*}
		\centering
		\includegraphics[width=\linewidth]{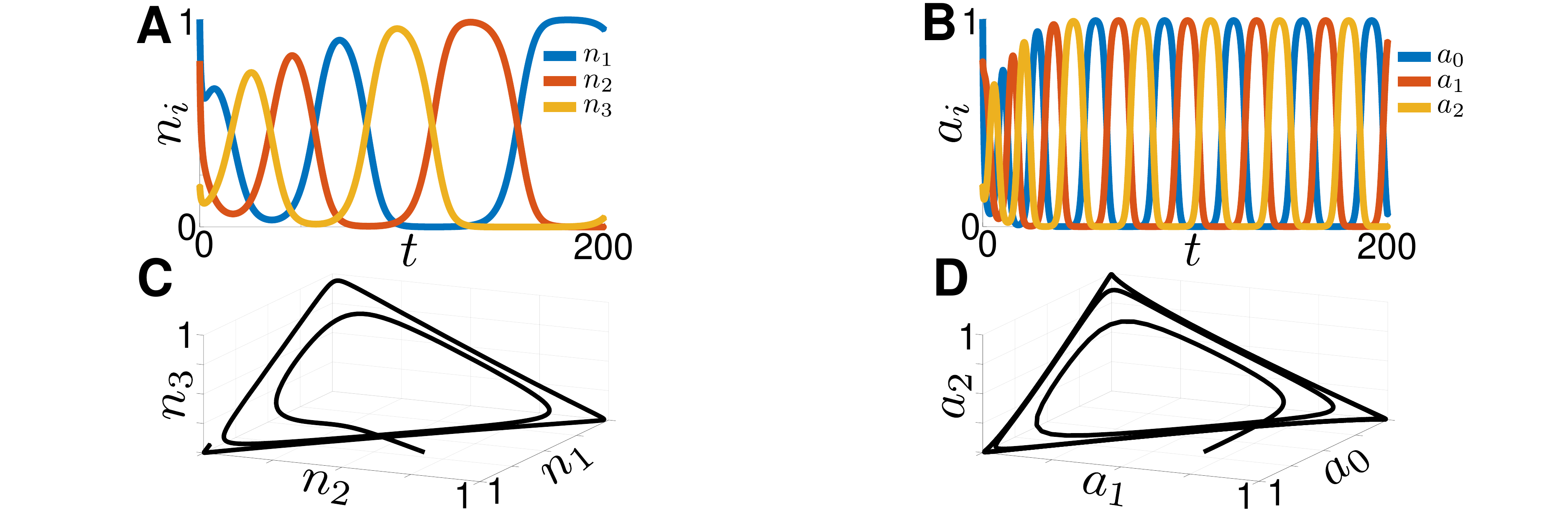}
		\caption{Comparison of deterministic minimal and three-pool models.
			\textbf{A:} Relative population size $n_i$ as a function of $t$ generated from \Cref{eq:Minimal_Model_MF} with $r=1$, $\alpha=0.8$, $\beta=1.3$, and $n(0)=(1,0.8,0.2)$.
			\textbf{B:} Active fraction of neural pool $a_i$ as a function of $t$ from \Cref{eq:Three_Pool_Deterministic} with $\tau=1$, $\gamma=2.4$, and $\mu=10^{-5}$ with same initial condition as \textbf{A}.
			\textbf{C:} Solution from \textbf{A} plotted in phase space.
			\textbf{D:} Solution from \textbf{B} plotted in phase space.}
		\label{fig:1}
	\end{figure*}
	To better understand the dynamics three-pool model, compared to the more traditional translations of heteroclinic cycling, we simulated the deterministic \Cref{eq:Minimal_Model_MF} and \Cref{eq:Three_Pool_Deterministic}  (see \cref{fig:1}). 
	\cref{fig:1}\textbf{A} shows that the minimal model exhibits the same heteroclinic cycling as May and Leonard's original system.
	As expected, solutions  converge to the plane $n_1+n_2+n_3=1$ in phase space (\cref{fig:1}\textbf{C}).
	In contrast, in the three-pool model, when $\mu=0$ we recover a rescaled version of the original May-Leonard system.
	However, when $\mu>0$, the three-pool model does not exhibit heteroclinic cycling; instead, as  \cref{fig:1}\textbf{B} shows, it undergoes periodic oscillations.
	Trajectories no longer converge to the triangular unit plane, but instead converge to a hyperbolic manifold (see \cref{fig:1}\textbf{D}).
	
	In the following sections we introduce stochastic models corresponding to each of the three mean-field models discussed above.
	In order to constrain our choice of stochastic model, in each case we restrict consideration to models that obey mass-action kinetics.
	This choice allows us to leverage results from birth-death processes and complex-balanced equilibrium theory in order to study the long-time asymptotic behavior of each model.
	As a consequence of this modeling choice, the noise in our models will come from demographic stochasticity rather than, for example, scaled Gaussian noise typical of Langevin-type population models.
	By focusing on discrete-state population models, we aim to hew closely to the spirit of May and Leonard's original work.  
	%While this modeling choice excludes many mathematically viable systems from our study, it grounds our work in the physical reality of population dynamics and remains in the same general spirit as May and Leonard's original system.
	
	\section{Stationary Distribution of General Variance and Minimal Models}
	\label{sec:Stationary_Dist}
	\subsection{General Variance Model: Total Extinction}
	\label{sec:GV_Model}
\begin{figure*}
		\includegraphics[width=\linewidth]{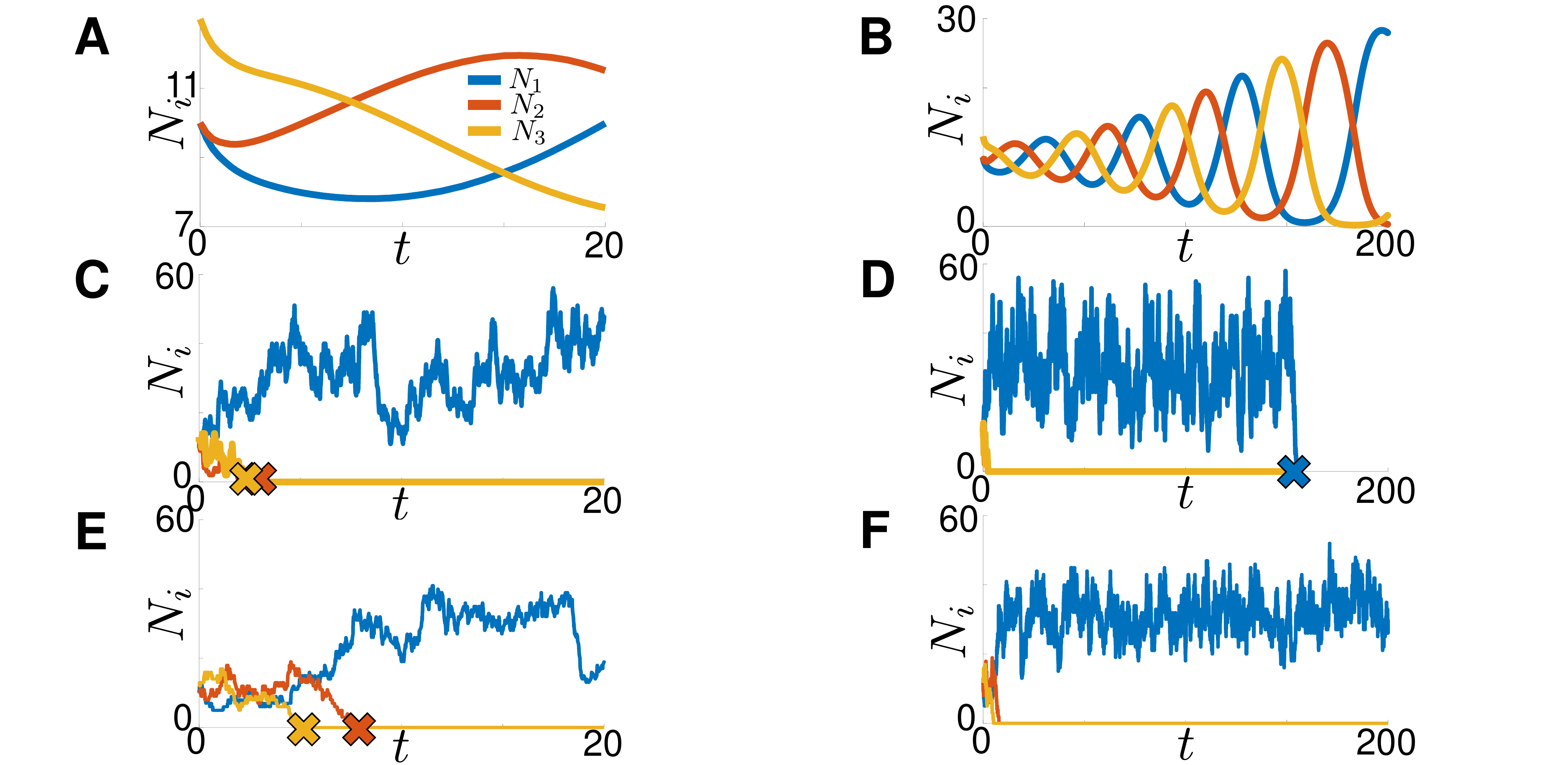}
		\caption{GV and minimal models of stochastic heteroclinic cycling. 
			\textbf{A:} Mean-field behavior of GV/minimal model from \Cref{eq:GV_Model_MF} with $\alpha = 0.8$, $\beta = 1.3$, $r=1$, $\Omega=30$, and $N(0)=\Omega\left(\frac{1}{3},\frac{1}{3},\frac{13}{30}\right)$. 
			\textbf{B:} Same as \textbf{A}, plotted for $0\le t \le 200$.  
			\textbf{C:} Stochastic realization of GV model from reaction system in \Cref{eq:GV_Model_Reactions_1}-\cref{eq:GV_Model_Reactions_5} with same parameters as  \textbf{A}.
			Strikes (\textbf{x}) mark extinction events of each species.
			\textbf{D:} Same as \textbf{B}, plotted for $0\le t \le 200$.
			Note that all species have gone extinct by the end of the simulation.
			\textbf{E:} \textbf{C:} Stochastic realization of minimal model from reaction system in \Cref{eq:Minimal_Model_Reactions_1}-\cref{eq:Minimal_Model_Reactions_4} with same parameters as \textbf{A}.
			\textbf{F:} Same as \textbf{E}, plotted for $0\le t \le 200$.
			In the minimal model, the last species survives in perpetuity.}
		\label{fig:2}
	\end{figure*}

	Following \cite{Higham08,wilkinson18}, we adopt the formalism of stochastic mass-action kinetics and construct the reaction net for \Cref{eq:GV_Model_MF}:
	\begin{align}
		\label{eq:GV_Model_Reactions_1}
		N_i&\xrightarrow{c_1}2N_i  &&(c_1 = b),&& \text{birth}\\
		N_i&\xrightarrow{c_2}\emptyset &&(c_2 = d),&& \text{death}\\
		2N_i&\xrightarrow{c_3}N_i  &&\left(c_3 = \frac{2}{\Omega}\right),&&\text{homocidal competition} \\
		N_i+N_{i+1}&\xrightarrow{c_4}N_{i+1} &&\left(c_4 = \frac{\alpha}{\Omega}\right),&&\text{heterocidal competition} \\
		N_{i}+N_{i+2}&\xrightarrow{c_5}N_{i+2} &&\left(c_5 = \frac{\beta}{\Omega}\right).&&\text{heterocidal competition}
		\label{eq:GV_Model_Reactions_5}
	\end{align}
	In each of \Cref{eq:GV_Model_Reactions_1}-\cref{eq:GV_Model_Reactions_5} we take $i\in\{1,2,3\}$ and interpret indicial addition cyclically.
	For each reaction, $c_j$ is the microscopic rate constant determining the propensity of the given reaction.
	\cref{fig:2}\textbf{B},\textbf{D} shows a sample trajectory of this system generated via Gillespie's stochastic simulation algorithm \cite{gillespie1977,Higham08}.
	While short-term dynamics of the mean-field model \Cref{eq:GV_Model_MF} evolve slowly from initial conditions (\cref{fig:2}\textbf{A},\textbf{B}),  the stochastic GV system quickly exhibits extinction of two of the three species (\cref{fig:2}\textbf{C}).
	This result is consistent with May and Leonard's prediction: while intensive variables can become infinitely close to zero, extensive variables taking discrete values will eventually drop to zero.
	However, rather than leading the third ``winning" species to dominate in perpetuity, on a longer time scale (\cref{fig:2}\textbf{D}) the winning population also ultimately suffers a downward fluctuation leading to its own extinction.  
	Indeed, the following proposition establishes that the unique stationary distribution for the general variance stochastic model is total extinction.  
	\begin{proposition}
		\label{prop:GV_Model}
		Let $\mathbf{N}=(N_1,N_2,N_3)$ be the vector of individuals in each population of the GV model. 
		If the per capita death rate $d>0$, then the unique stationary distribution of the reaction system \Cref{eq:GV_Model_Reactions_1}-\cref{eq:GV_Model_Reactions_5} is $\pi(\mathbf{n})=\delta(\mathbf{n})$.
	\end{proposition}
	We provide a proof in \cref{app:GV_Model}.
	
	\cref{prop:GV_Model} tells us that the GV model will exhibit total extinction in the long-time limit, independent of initial conditions.
	This result recalls that of Vellela and Qian \cite{Vellela2007}, in which the authors demonstrated that for the single-population Keizerator reaction system, the mean-field system converges to a nontrivial equilibrium while the stochastic system converges to total extinction (albeit with  mean extinction times that are exponentially long in the system size).
	We can explain this behavior by the fact that our reaction system includes individual birth ($N_i\to 2N_i$) and death ($N_i\to\emptyset$) reactions \cite{allen2010}.
	As we will see in \cref{sec:Minimal_Model}, removing the death reaction fundamentally changes the stationary behavior of the model.
	
	\subsection{Minimal Model: Persistence of a Single Species}
	\label{sec:Minimal_Model}
\begin{figure*}
		\centering
		\includegraphics[width=\linewidth]{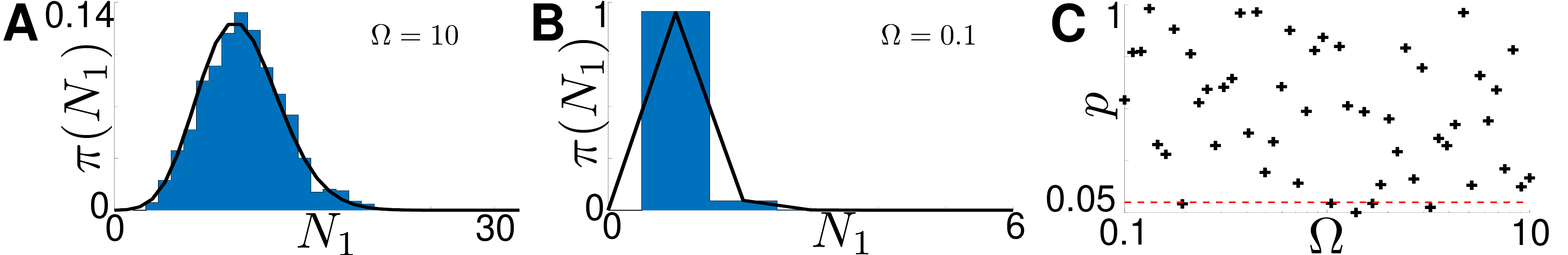}
		\caption{Comparison of analytical and empirical stationary distributions for the minimal variance model.
			\textbf{A:} Blue bars show histogram of empirical stationary distribution from Gillespie simulations with $\alpha=0.8$, $\beta=1.3$, $r=1$, and $\Omega=10$.
			Black curve shows stationary distribution calculated from \Cref{eq: Minimal Model Stationary Distribution} with $\Omega=10$. 
			\textbf{B:} Same as \textbf{A}, with $\Omega=0.1$.
			\textbf{C:} $p$-values from $\chi^2$ goodness-of-fit test between empirical stationary distribution from Gillespie simulations and analytical stationary distribution from \Cref{eq: Minimal Model Stationary Distribution} with varied $\Omega$.
			Dashed line shows standard significance level $\alpha=0.05$; $p$-values above the dashed line indicate good agreement between the distributions.}
		\label{fig:3}
	\end{figure*}

	We previously noted that \Cref{eq:Minimal_Model_MF} is a special case of \Cref{eq:GV_Model_MF}, obtained by setting $d=0$ and $b=r$.
	Setting $d=0$ is equivalent to removing the individual death reactions $N_i\to \emptyset$, thus the  reaction system takes the form (for $i\in\{1,2,3\}$, as before):
	\begin{align}
		\label{eq:Minimal_Model_Reactions_1}
		N_i&\xrightarrow{c_1}2N_i && (c_1 = r), &&\text{birth}\\
		2N_i&\xrightarrow{c_2}N_i && \left(c_2 = \frac{2}{\Omega}\right), &&\text{homocidal competition}\\
		N_i+N_{i+1}&\xrightarrow{c_3}N_{i+1} && \left(c_3 = \frac{\alpha}{\Omega}\right),&&\text{heterocidal competition} \\
		N_i+N_{i+2}&\xrightarrow{c_4}N_{i+2} && \left(c_4 = \frac{\beta}{\Omega}\right) && \text{heterocidal competition}.            
		\label{eq:Minimal_Model_Reactions_4}
	\end{align}
	\cref{fig:2}\textbf{E},\textbf{F} contrast Gillespie simulations of the GV model and the minimal model.  
	In both case two population extinctions occur quickly, but in the minimal model the third population does not go extinct.
	In the minimal model, once the system reduces to a single species, the only death mechanism is homocidal competition, which requires at least two individuals.
	Therefore the total extinction state is not accessible from non-trivial initial conditions, which  guarantees a distinct stationary distribution from the GV model.
	The framework of complex-balanced equilibria from Horn and Jackson \cite{Horn1972} and Anderson and Kurtz \cite{Anderson2015a} allows us to obtain this stationary distribution for the system, as given in \cref{prop:minimal_model}:
	\begin{proposition}
		\label{prop:minimal_model}
		Let $\mathbf{N}=(N_1,N_2,N_3)$ be the population vector of the minimal model \Cref{eq:Minimal_Model_Reactions_1}-\cref{eq:Minimal_Model_Reactions_4}. 
		The reaction system \Cref{eq:Minimal_Model_Reactions_1}-\cref{eq:Minimal_Model_Reactions_4} has four distinct stationary distributions.
		Three may be expressed as component-wise stationary distributions of the form
		\begin{equation}
			\label{eq: Minimal Model Stationary Distribution}
			\pi(n_i)=\frac{\Omega^{n_i}}{n_i!(e^\Omega-1)}\delta(n_{i+1})\delta(n_{i+2}),
		\end{equation}
		for $i\in\{1,2,3\}$, $n_i\ge 1$, with $\delta(x)$ being the distribution with unit probability at $x=0$, and with index addition taken cyclically on $\{1,2,3\}$.
		The fourth is $\pi(\mathbf{n})=\delta(\mathbf{n})$.
	\end{proposition}
	We provide a proof in \cref{app:Minimal_Model}. 
	\cref{fig:3} shows a comparison of the analytic stationary distribution from \Cref{eq: Minimal Model Stationary Distribution} with the empirical distribution from Gillespie simulations.
	We can see that the two results show excellent agreement, even when the system size $\Omega$ takes on non-integer values.
	
	Comparing \cref{prop:GV_Model} and \cref{prop:minimal_model}, the stationary behaviors of the GV and minimal models are in fact distinct.
	While two of the populations  will go extinct in both models, the GV model converges to total extinction while the minimal model converges to a truncated Poisson distribution representing stochastic logistic growth.
	Although the total extinction state is a stationary distribution for the minimal model, it is not accessible from nontrivial initial conditions.
	This comparison illustrates the well-known fact that two stochastic models both consistent with the same mean-field deterministic model can have fundamentally different long-term behavior.
	
	\section{Transient Behavior of the Minimal Model}
	\label{sec:ML_Transient_Behavior}
	While we have thus far restricted our investigation to long-time asymptotic behaviors, we may also study the dynamics of extinction over intermediate times.
	The order and timing of extinctions is important in conservation ecology, where it is crucial to determine if and when intervention is required to prevent population collapse \cite{shaffer81,purvis2000}.
	Gillespie simulations suggest that the cycle length of the stochastic May-Leonard system, conditioned on non-extinction, has finite mean (\cref{fig:2}).
	Taking this observation together with the stationary distribution results from \cref{sec:Stationary_Dist}, we can reasonably expect to estimate both the ordering of extinction events and their times of the stochastic system. 
	Because both the GV and minimal models exhibit similar transient behavior, we will restrict our investigations to the minimal model, as the results will be more clear due to its lower variance.
	\subsection{Distribution and Ordering of Extinction Events}
	\label{sec:Minimal_Extinction_Distribution}
\begin{figure*}
		\includegraphics[width=\linewidth]{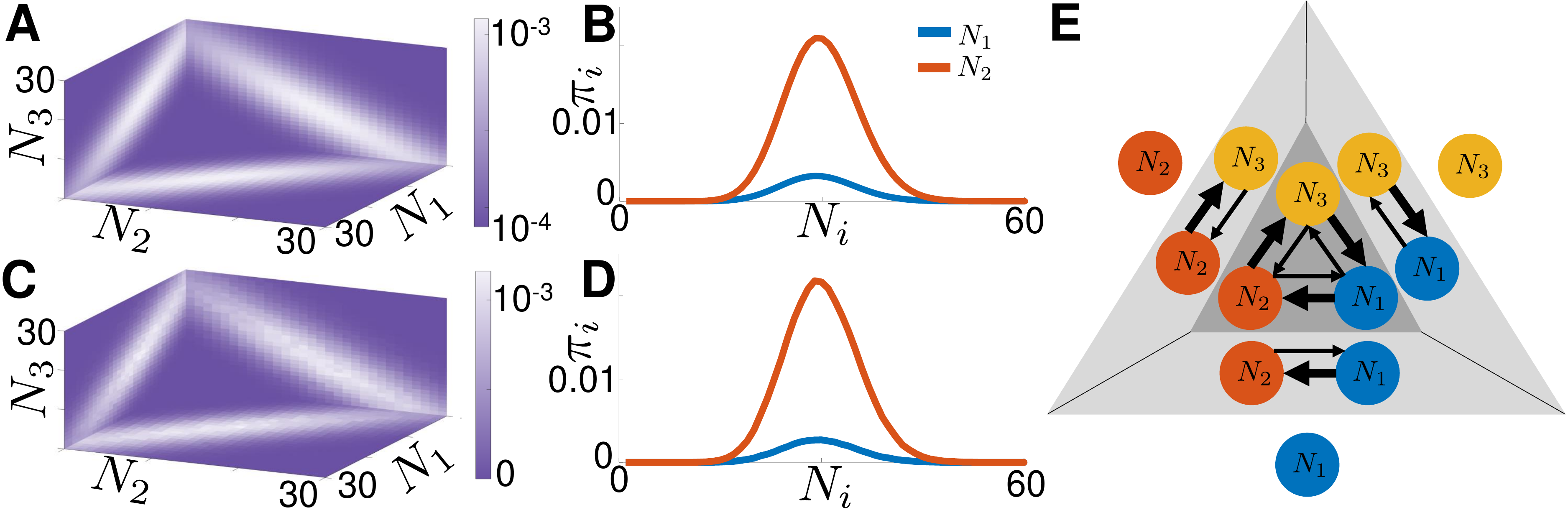}
		\caption{Ordering of extinction events in the minimal-variance model.
			\textbf{A:} Exact distribution for first extinction event obtained by solving the discrete first-passage location problem ($r=1$, $\alpha=0.8$, $\beta = 1.3$, $\Omega=30$). 
			\textbf{B:} Exact distribution for second extinction event, conditioned on $N_3$ going extinct first, obtained by solving the discrete first-passage location problem.
			Red (upper) curve: Density of population $N_1$ when population $N_2$ goes extinct. 
			Blue (lower) curve: Density of population $N_2$ when population $N_1$ goes extinct. 
			See text for details.                   
			\textbf{C:} Empirical distributions for first extinction event, obtained from $10^6$ Gillespie simulations with the same parameters as in \textbf{A}. 
			\textbf{D:} Empirical distributions for second extinction event, conditioned on $N_3$ going extinct first, obtained from Gillespie simulations with $10^6$ initial states sampled from the plane $N_3=0$ in \textbf{C}. 
			\textbf{E:} Schematic showing competition interactions and ordering of extinction events. Thicker arrows indicate stronger  competitive interactions.}
		\label{fig:4}   
	\end{figure*}

	To study the ordering of extinction events in the minimal model, we found the distribution of hitting locations on the coordinate planes $N_i=0$ for $i\in\{1,2,3\}$ from a fixed initial condition $\mathbf{N}(0)=\frac{\Omega}{3}(1,1,1)$.
	This distribution gives the relative probability of extinction of each species from this starting condition. 
	In addition, it gives us the conditional density of, for example, species 2 and 3, conditioned on species 1 going extinct first. 
	We formulated the first-passage location problem as
	\begin{equation}
		\mathcal{L}\pi=\mathbf{e}_s,
		\label{eq:first-passage_location}
	\end{equation}
	where $\mathcal{L}$ is the infinitesimal generator matrix associated with the discrete master equation, $s$ is a fixed absorbing state,  $\pi$ is the probability of hitting $s$ as a function of initial condition, and $\mathbf{e}_s$ is the standard basis vector.
	We imposed absorbing boundary conditions along the coordinate planes $N_i=0$ and adjoint reflecting boundary conditions along the planes $N_i=2\Omega$ to ensure a well-posed numerical problem in which probability conservation is guaranteed.
	For more details about this construction, see \cref{app:Discrete_Operator}.
	\cref{fig:4}\textbf{A} shows the solution of this linear system; we can see that for an initial condition along the vector $(1,1,1)$, the distribution of absorption locations exhibits a three-fold  rotational symmetry about the initial condition.  
	The majority of the distribution is located near the intersections of the plane $N_1+N_2+N_3=\Omega$ with the three absorbing coordinate planes.
	These results suggest that, for a symmetrically-distributed initial condition, all three populations are equally likely to go extinct.
	To confirm these findings, we also found the first-hitting distribution empirically, shown in \cref{fig:4}\textbf{C}, using Gillespie simulations.
	We can see that the two approaches show good agreement over the entire domain.
	
	Assuming WLOG that $N_3$ is the first population to go extinct, we formulated the first-hitting problem for the two-dimensional subsystem to find the absorption distribution of the remaining two species conditioned on extinction of $N_3$.
	Using the same approach from the full three-dimensional system, with initial conditions weighted by the distribution in \cref{fig:4}\textbf{A} over the plane $N_3=0$, we obtained the distributions shown in \cref{fig:4}\textbf{B}.
	The upper (red) curve labeled ``$N_2$'' shows the density of $N_1$ at the time $N_2$ goes extinct. 
	Similarly, the lower (blue) curve labeled ``$N_1$'' shows the density of $N_2$ at the time $N_1$ goes extinct.
	The area under each curve gives the conditional probability that the corresponding population goes extinct, given that $N_3$ goes extinct first; note that the summed area under the two curves equals unity.
	From these results, we can see that once $N_3$ goes extinct, $N_2$ is much more likely to go extinct than $N_1$.
	We again confirmed our these results using Gillespie simulations (\cref{fig:4}\textbf{D}) and found good agreement ($\chi^2$ goodness-of-fit test, $p=0.96>0.05$).
	
	Combining these two results, we can see that there is a distinct pattern to the extinctions in the minimal model, which is schematized in \cref{fig:4}\textbf{E}.
	In the full three-dimensional system, the likelihood of each extinction is determined by the initial conditions; any initial condition along the vector $(1,1,1)$ results in an equal probability of first extinction.
	Once one population goes extinct, a second population quickly goes extinct because of the imbalance in competition rates $\alpha$ and $\beta$, leaving a sole surviving species.
	For example, if species 3 goes extinct first, then it is more likely that species 2 goes extinct next, leaving species 1 to dominate over long times.
	This pattern is reminiscent of the age-old saying ``the enemy of my enemy is my friend,'' as species 1, which is out-competed by species 3, survives because species 2 out-competes species 3.
	\subsection{Extinction Times in the Minimal Model}
	\label{sec:Minimal_Extinction_Times}
\begin{figure*}
		\includegraphics[width=\linewidth]{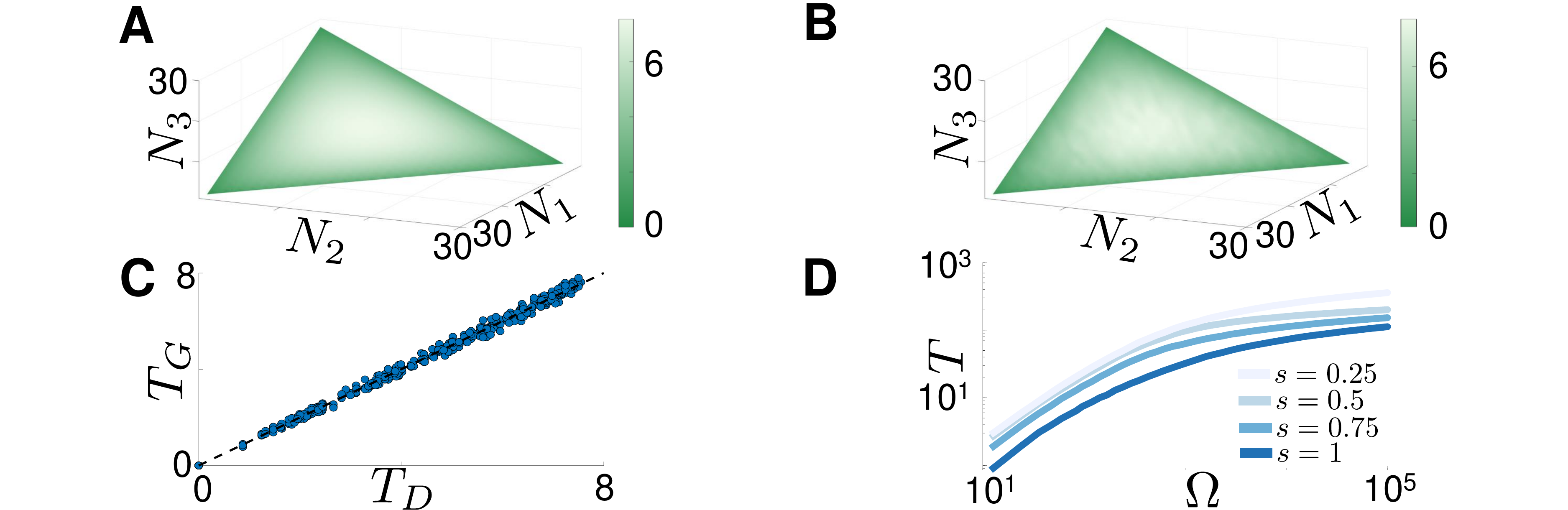}
		\caption{Timing of first extinction in the minimal model.
			\textbf{A:} Exact mean time to first extinction event, obtained by solving the discrete first-passage time problem ($r = 1$, $\alpha=0.8$, $\beta = 1.3$, $\Omega=30$), from initial conditions in the plane $\Pi=\{N_1+N_2+N_3=\Omega/3\}$.
			\textbf{B:} Empirical expected time to first extinction event, calculated using $10^4$ Gillespie simulations over $\Pi$ with the same parameters as in \textbf{A}.
			\textbf{C:} Scatter plot of mean extinction times at every initial condition.
			Abscissa: Exact time $T_D$, from discrete backward equation. Ordinate: Empirical mean time $T_G$, from $10^4$ samples.
			\textbf{D:} Empirical mean time to first extinction event, obtained from $10^4$ Gillespie simulations, as a function of $\Omega$.
			Several values of $s$ along the segment $\left[\Omega(1-s)+\frac{\Omega s}{3},\frac{\Omega s}{3},\frac{\Omega s}{3}\right]$ are superimposed.}
		\label{fig:5}
	\end{figure*}

	%    The first-passage location distribution of the full three-dimensional system (\cref{fig:4}) suggests that once a first population goes extinct, a second extinction will follow in quick succession.
	%    The second extinction is expected to occur quickly because the first extinction usually leaves one surviving population much higher than the other.
	%    This observation mirrors what we previously saw in Gillespie simulations, and leads us to restrict our focus to finding the expected time of the first extinction event.
	In order to find the exact mean time to first extinction, we construct the first-passage time problem
	\begin{equation} 
		\mathcal{L}\mathbf{\tau}=\mathbf{-1},
		\label{eq:first-passage_time}
	\end{equation}
	where $\mathcal{L}$ is the same infinitesimal generator matrix from \Cref{eq:first-passage_location}, $\tau$ is the vector of mean absorption times as a function of initial condition, and $\mathbf{-1}$ is a vector of -1's.
	We impose absorbing boundary conditions on the coordinate planes $N_i=0$, and adjoint reflecting boundary conditions on the planes $N_i=2\Omega$, as in the first-passage location problem (\cref{sec:Minimal_Extinction_Distribution}).
	Using this approach, we obtained the mean first-extinction time for all initial states in the domain.  
	For ease of visualization, \cref{fig:5}A and B show $\tau$ restricted to 
	%, we found that initial conditions taken off
	the plane $\Pi=\{N_1+N_2+N_3=\frac{\Omega}{3}\}$.
	(We note that trajectories with initial conditions away from $\Pi$ quickly approach a small neighborhood of this plane, so mean extinction times on the plane are representative of mean extinction times from most starting locations in the interior of the domain.)
	\cref{fig:5}\textbf{A} shows a slice of the mean first-extinction time along the plane $\Pi$.
	As expected, the extinction times as a function of initial condition have three-fold rotational symmetry. 
	Moreover, the time is largely determined by the distance between the initial condition and the deterministic fixed point $\frac{\Omega}{3}(1,1,1)$.
	
	To confirm the results from the discrete first-passage time problem, we also used large-sample Gillespie simulations with initial condition taken over $\Pi$.
	\cref{fig:5}\textbf{B} shows the empirical mean first-extinction time as a function of starting location.
	Comparing the exact and approximate results, we found good agreement ($t$-test, averaged over initial conditions: $\langle p(\mathbf{N})\rangle_{\Pi}\approx0.51>0.05$).
	To illustrate  this agreement further, in \cref{fig:5}\textbf{C} we plot each initial condition in a scatter plot: the abscissa is the exact mean extinction time found using the discrete backward equations ($T_D$) and the ordinate is the empirical mean extinction time found using Gillespie simulations ($T_G$).
	The two methods show excellent agreement for initial states near the coordinate planes (when $T_D$ is small) and have a slightly increased variance when the initial state is close to the deterministic fixed-point (when $T_D$ is large).  
	Nevertheless $T_D$ and $T_G$ show excellent agreement over the entirety of $\Pi$.
	This result demonstrates that large-sample Gillespie simulations give a good approximation of the exact mean extinction times, and justifies the use of Gillespie simulations for large-$\Omega$ systems where the exact solution becomes intractable (e.g.~$\Omega\gtrsim 60$).
	
	In order to study the effect of system size  on mean extinction time for $\Omega>30$, we relied on Gillespie simulations.
	From our previous simulations for a fixed $\Omega=30$ system, we observed that time to first extinction has three-fold rotational symmetry and largely depends on the distance from the deterministic fixed point.
	Therefore we considered initial conditions along the segment connecting $(\Omega,0,0)$ and $\frac{\Omega}{3}(1,1,1)$, where we parameterized the distance along this segment using the parameter $s\in[0,1]$.
	Using this parameterization, we varied $\Omega$ and $s$ and estimated the time to first extinction, shown in \cref{fig:5}\textbf{D}.
	As $s$ increases and the initial condition moves closer to the deterministic fixed point, the mean extinction time increases across all values of $\Omega$; this trend is consistent with the behavior we observe in the small-$\Omega$ system.
	
	\section{Three-Pool Model: Stochastic Oscillations}
	\label{sec:Three_Pool_Model}
	
	In the previous sections, we showed that modifying a single reaction in the stochastic model (removing the individual death reaction) led to distinct asymptotic dynamics.
	However, both the GV and minimal models share the same mean-field behavior, and both produce transient dynamics that may be described as noisy heteroclinic cycling.
	In contrast, the three-pool model for a neuromotor central pattern generator (CPG) in \Cref{eq:Three_Pool_Deterministic} has a non-homogeneous term, $\mu$, that steers trajectories away from the fixed points in the corners of the boundaries, preventing heteroclinic cycling.
	In the CPG model, the parameter $\mu$ represents endogenous activation of each pool of motor neurons.
	When $\mu>0$
	the resulting deterministic system exhibits finite-period oscillations, converting heteroclinic cycling into finite-period limit cycle behavior (see \cref{fig:1}), with prolonged dwell times near the saddle points and a period that can be sensitively controlled by the endogenous activation parameter.
	
	Both endogenous activation and noise intensity have been suggested as potential  mechanisms for regulating the frequency of cycling in CPG models built on a dynamical architecture of heteroclinic cycling \cite{Shaw2014PhDThesis,shaw2012,shaw15,Lyttle2017}.       
	The three-pool model specified below allows us to investigate the relative contributions of both activation (controlled by $\mu$) and noise (controlled by the system size $\Omega$) to regulating the mean oscillation period of the CPG model.
	
	Using the same formalism as in \cref{sec:Stationary_Dist}, we write the reaction net for $A_i$ for the three-pool system as:
	\begin{align}
		\label{eq:Three_Pool_Reactions_1}
		A_i+I_i&\xrightarrow{c_1}2A_i && \left(c_1 = \frac{1}{\tau\Omega}\right), &&\text{self-activation}\\
		A_i+A_{i+1}&\xrightarrow{c_2}I_i+A_{i+1} && \left(c_2 = \frac{\gamma}{\tau\Omega}\right), &&\text{inhibition of }i\text{ by }i+1\\
		I_i&\xrightarrow{c_3}A_i && \left(c_3 = \frac{\mu}{\tau}\right),&&\text{endogenous activation}
		\label{eq:Three_Pool_Reactions_3}
	\end{align}
	where $i\in\{0,1,2\}$ and indicial addition is taken cyclically; recall that $I_i=\Omega-A_i$ is the inactive population.
	Note that the endogenous activation $\mu$ enters into the reaction $I_i\to A_i$.
	Because the total population of cells in each pool remains fixed over time, this reaction ensures that even if one population becomes fully inactive, it will eventually become active again, after some delay.
	Thus, in the language of the previous two models, the neural populations in the three-pool model will never go permanently extinct.
	Consequently the neural activity oscillation persists indefinitely, albeit with a randomly varying cycle length.
	
\begin{figure*}
		\centering
		\includegraphics[width=\linewidth]{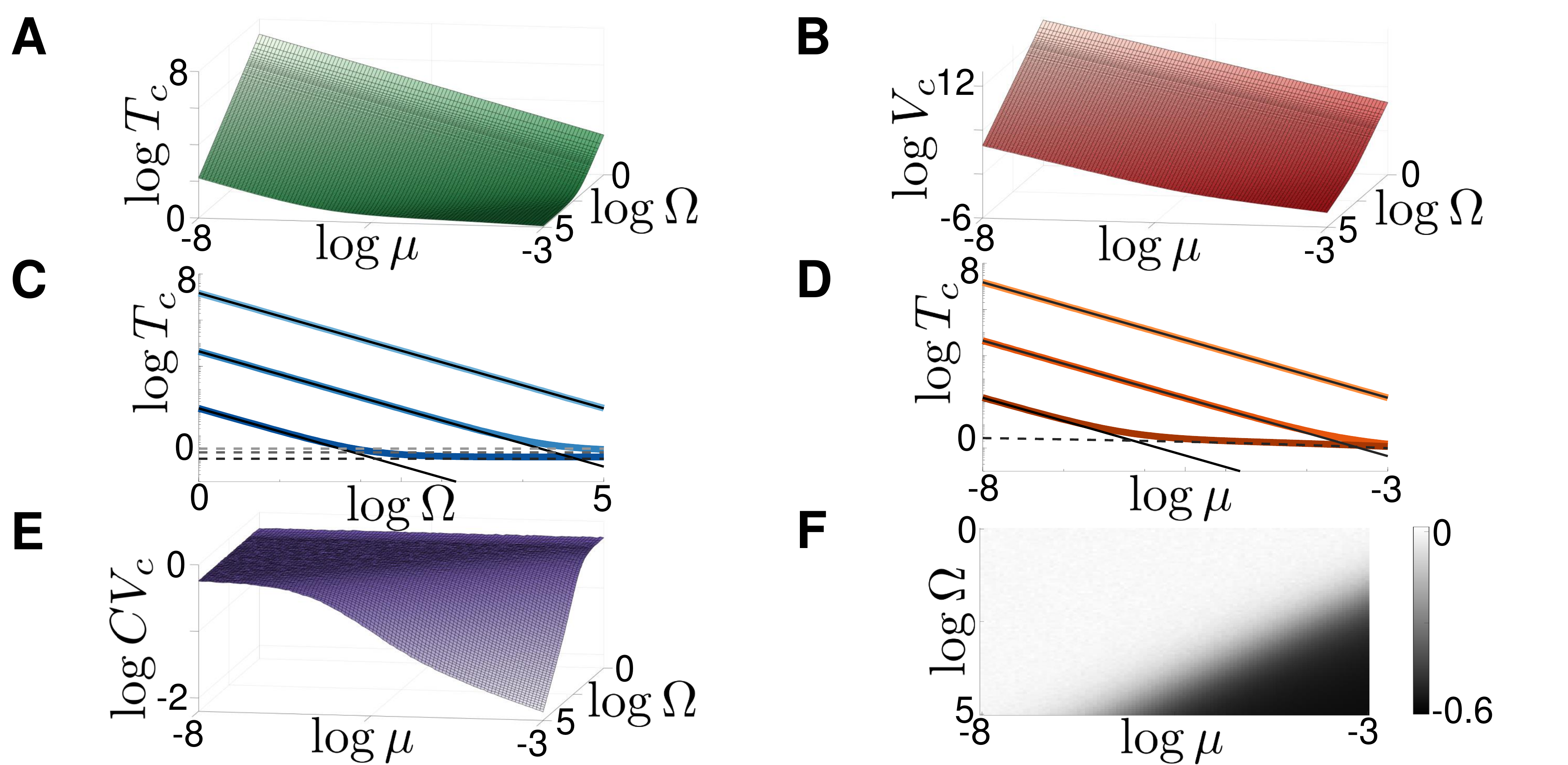}
		\caption{Cycle length statistics for the three-pool model.
			\textbf{A:} Mean cycle length $T_c$ as a function of system size $\Omega$ and excitation parameter $\mu$. 
			Mean calculated at each parameter set using $10^4$ samples from Gillespie simulations.
			\textbf{B:} Variance in cycle length as a function of $\Omega$ and $\mu$.
			\textbf{C:} Mean cycle length as a function of $\Omega$, with several values of $\mu$ superimposed.
			Solid thin lines show the approximate mean cycle length given by \Cref{eq:Three_Pool_Approximate_Cycle_Length}, and dashed lines show cycle length of deterministic system from \Cref{eq:Three_Pool_Deterministic} for the given value of $\mu$.
			\textbf{D:} Mean cycle length as a function of $\mu$, with several values of $\Omega$ superimposed.
			Solid thin lines show approximate cycle length given by \Cref{eq:Three_Pool_Approximate_Cycle_Length}, and dashed line shows deterministic cycle length as a function of $\mu$.
			\textbf{E:} Coefficient of variation (CV) of the cycle length as a function of $\Omega$ and $\mu$.
			\textbf{F:} Difference in CV from Gillespie samples $(CV_c)$ and CV from Gamma distribution approximation $(CV_a)$, calculated as $\Delta CV=CV_c-CV_a$, as a function of $\Omega$ and $\mu$. 
		}
		\label{fig:6}
	\end{figure*}

	\cref{fig:6} illustrates how the population size $\Omega$ and activation strength $\mu$  influence the cycle length.
	In order to cover a wide range of system sizes, we utilized Gillespie simulations.  
	\cref{fig:6}\textbf{A} shows the empirical mean cycle length for varied $\Omega$ and $\mu$; we observe that larger parameter values cause faster oscillations, on average. 
	Additionally, as $\Omega$ increases, the mean period approaches a value that depends solely on $\mu$; this value is the deterministic period from the mean-field equations in \Cref{eq:Three_Pool_Deterministic}.
	We  calculated the empirical variance of the cycle length, shown in \cref{fig:6}\textbf{B}, and found that the variance also decreases when either $\Omega$ or $\mu$ are increased.
	These results suggest that both $\mu$ and $\Omega$ could contribute to controlling the frequency of neural activity.
	For example, consider a relatively slow system, with small $\Omega$ and small $\mu$.
	This system can be sped up by either increasing $\mu$, which increases endogenous activation noise and drives activity further away from the saddle points, or by increasing $\Omega$, which decreases demographic stochasticity.
	Additionally, both parameters have similar influence on the variance of the cycle length.
	Recent work has shown that the feeding CPG of the marine mollusk \textit{Aplysia californica} recruits additional motor neurons when the organism encounters unexpected resistance in swallowing food \cite{gill2020,gill2020thesis}, and that variability of motor neuronal activity is reduced for those components of feeding behavior that matter most for task fitness \cite{cullins2015}.
	Although the isolated three pool model considered here lacks important circuit components (such as sensory feedback \cite{cullins2015b}), the relative sensitivity of the cycle time variance to $\mu$ versus $\Omega$ could nevertheless suggest experimentally testable questions.   
	For instance, one could probe experimentally whether the variability in the motor pattern decreases or increases when subjected to larger external loads.
	%Our model accomplishes this additional recruitment by increasing $\Omega$, which would predict both faster average neural activity and lower variance.
	%This testable hypothesis suggests the organism is able to exert more fine motor control by modulating noise.
	
	\cref{fig:6}\textbf{A},\textbf{B} exhibit a large region of parameter space in which the mean and variance both vary linearly on a log scale with both $\Omega$ and $\mu$.
	To explain this observation, we developed an approximate expression for the average period as a function of $\Omega$ and $\mu$.
	Consider the discrete system, which forms a cubic lattice, and suppose the population vector is currently $(\Omega,0,0)$.
	While there are three possible transitions away from this state, the only transition that pushes the system forward along a cycle is the transition  $(\Omega,0,0) \to (\Omega,1,0)$.
	The time of this transition is exponentially distributed with rate parameter $\frac{\Omega\mu}{\tau}$.
	Once this transition occurs, the subsequent transitions are more rapid, and push the system to the corner $(0,\Omega,0)$, where the process repeats itself.
	Because of the differing timescales of these transitions, we can approximate the cycle dynamics as a sequence of three ``rate-limiting steps'', each with transition times that are iid exponentially distributed with parameter $\frac{\Omega\mu}{\tau}$. 
	The sum of these three times follows a Gamma distribution with parameters $(\alpha,\beta)=\left(3,\frac{\tau}{\mu\Omega}\right)$.
	This distribution predicts the mean cycle length and coefficient of variation to be
	\begin{equation}
		T_a\approx \frac{3\tau}{\Omega\mu},\quad\quad \text{CV}_a\approx\frac{1}{\sqrt{3}}.
		\label{eq:Three_Pool_Approximate_Cycle_Length}
	\end{equation}
	
	To verify the Gamma distribution approximation and the predicted mean cycle length given by \Cref{eq:Three_Pool_Approximate_Cycle_Length}, we plot the average period as a function of $\Omega$ for fixed $\mu$, and as a function of $\mu$ for fixed $\Omega$ (solid thin lines) against the empirical average period (thick colored lines) in \cref{fig:6}\textbf{C} and \cref{fig:6}\textbf{D}, respectively. 
	The Gamma distribution and Gillespie simulation results agree in the region of small $\Omega$ and small $\mu$; however, the Gamma distribution approximation breaks down as $\Omega$ increases and the number of different transition paths between fixed points increases.
	To further validate the Gamma distribution approximation, in \cref{fig:6}\textbf{E} we plot the coefficient of variation (CV) of the empirical cycle length and found that the CV is approximately constant for a large region of parameter space.
	Comparing our empirical CV to the predicted value \Cref{eq:Three_Pool_Approximate_Cycle_Length} in \cref{fig:6}\textbf{F}, we found that this heuristic interpretation holds for a large range of parameters.
	
	\section{Conclusions}
	\label{sec:conclusions}
	In this work, we studied both transient and long-term behavior in several stochastic versions of May and Leonard's heteroclinic cycling model, introducing noise via demographic stochasticity under a variety of assumption.
	%In this work, we constructed several stochastic implementations of heteroclinic cycling and studied their different long-term behaviors.
	Although two of our models (the general-variance (GV) and minimal models) coincide with the classical May-Leonard system in \Cref{eq:HC_Deterministic} in the mean-field limit, we found that these stochastic versions are guaranteed to undergo population extinctions in finite time.
	Moreover, by eliminating the individual death reactions in the GV model to obtain the minimal model, we proved that the stationary distribution changes from total extinction of all species (in the GV model) to extinction to a sole survivor that follows a truncated Poisson distribution (in the minimal model).
	We also studied a variant of the model representing a three-pool neural system.  
	In this version, we added a single reaction to introduce endogenous excitation of each neural population; in an ecological context a similar modification can be thought of as representing immigration.
	This additional reaction yielded a system that not only avoids permanent extinctions, but  has a finite mean cycle time that depends on the size of each neural pool ($\Omega$) and the strength of endogenous excitation ($\mu$).
	Using an intuitive rate-limiting step argument, we found an approximation to the mean cycle length that showed good agreement in both mean and variance with Gillespie simulations.
	As elements of a potential control scheme for a neural central pattern generator, it is worth noting that although both $\Omega$ and $\mu$ provide potential control parameters, their effects on the mean and variance of the cycle time are similar for a wide range of parameters, meaning that the mean and variance cannot be controlled independently of one another.
	
	Throughout our investigation, we limited attention to stochastic models with mass-action kinetics.
	This focus allowed us to formulate our models as multi-dimensional birth-death processes and leverage results from complex balanced equilibrium theory to find the stationary distribution of the GV and minimal models.
	While others, such as Reichenbach et.~al.~\cite{Reichenbach2006} and Yahalom et.~al.~\cite{yahalom2019}, have studied cyclic stochastic population models, their results required taking continuum limits of the state space and linearizing the resulting dynamics about fixed-points of the associated deterministic system.
	As a result, their models employed Gaussian white noise rather than discrete population noise, which can lead to inconsistent treatment of small population dynamics leading to extinction \cite{Strang2019}.
	Our approach avoided these potential difficulties and guaranteed that demographic stochasticity was the only source of noise in our models.
	
	To verify our analytic results, we both ran large-scale Gillespie simulations and constructed first-passage problems as sparse linear systems.
	While these two numerical approaches showed good agreement, we could only leverage the exact results from the master equation for small system sizes.
	This limitation is a consequence of our choice to use a birth-death formalism for all the stochastic models: the infinitesimal generator matrix $\mathcal{L}$ that is required for solving first-passage problems has $\mathcal{O}\left(\Omega^3\right)$ scaling, and we quickly reached hardware limitations when trying to vary $\Omega$ over several orders of magnitude.
	Future work using the discrete system may require approximating the operator to make the first passage problem tractable.
	Safta et.~al.~\cite{safta2015} have developed a hybrid discrete-continuum approximation of the forward operator, the adjoint of the infinitesimal generator matrix, that involves partitioning the state space, taking a continuum limit within each partition, and simulating a continuous flow within partitions and discrete transitions between partitions.
	In future work, this approach could be extended to create a hybrid approximation to the backward operator $\mathcal{L}$ to solve first-passage time problems, which would allow us to obtain the extinction time statistics for a larger range of system sizes semi-analytically.
	
	\appendix
	\section{Proof of Proposition 3.1}
	\label{app:GV_Model}
	For the reader's convenience, we restate \cref{prop:GV_Model}:
	
	\textit{Let $\mathbf{N}=(N_1,N_2,N_3)$ be the vector of individuals in each population of the GV model. 
		If the per capita death rate $d>0$, then the unique stationary distribution of the reaction system \Cref{eq:GV_Model_Reactions_1}-\cref{eq:GV_Model_Reactions_5} is $\pi(\mathbf{n})=\delta(\mathbf{n})$.}
	
	Our proof follows ideas similar to Vellela and Qian \cite{Vellela2007}.
	\begin{proof}
		Following standard arguments \citep{taylor98,wilkinson18,Higham08,calvetti2012}, 
		we define the probability distribution $p_{i,j,k}(t)=\Pr\left(\mathbf{n}(t)=(i,j,k)\right)$, where $i,j,k\in\{0,1,2,\dots\}$.
		(We set $p_{i,j,k}\equiv 0$ if $i<0$, $j<0$ or $k<0$.) 
		Recall from the definition of the GV model (\Cref{eq:GV_Model_Reactions_1}-\cref{eq:GV_Model_Reactions_5}) that $b,\Omega>0$ and  $\alpha,\beta\ge 0$.
		The distribution $p_{i,j,k}$ obeys an evolution equation, or discrete master equation, of the form
		\begin{multline}
			\frac{dp_{i,j,k}}{dt} = b\left[(i-1)p_{i-1,j,k}+(j-1)p_{i,j-1,k}+(k-1)p_{i,j,k-1}\right]\\
			+d\left[(i+1)p_{i+1,j,k}+(j+1)p_{i,j+1,k}+(k+1)p_{i,j,k+1}\right]\\
			+(i+1)\left(\frac{i}{\Omega}+\frac{\alpha}{\Omega}j+\frac{\beta}{\Omega}k\right)p_{i+1,j,k}\quad
			+\quad (j+1)\left(\frac{\beta}{\Omega}i+\frac{j}{\Omega}+\frac{\alpha}{\Omega}k\right)p_{i,j+1,k}\\
			+(k+1)\left(\frac{\alpha}{\Omega}i+\frac{\beta}{\Omega}j+\frac{k}{\Omega}\right)p_{i,j,k+1}\quad
			-\quad (b+d)(i+j+k)p_{i,j,k}\\
			-\left[i\left(\frac{i-1}{\Omega}+\frac{\alpha}{\Omega}j+\frac{\beta}{\Omega}k\right)+j\left(\frac{\beta}{\Omega}i+\frac{j-1}{\Omega}+\frac{\alpha}{\Omega}\right)+k\left(\frac{\alpha}{\beta}i+\frac{\beta}{\Omega}j+\frac{k-1}{\Omega}\right)\right]p_{i,j,k}.
			\label{eq:GV_Model_DME}
		\end{multline}

		\textbf{Existence}: Setting $p_{0,0,0}=1$ and all other $p_{i,j,k}=0$ satisfies the equilibrium condition $\frac{dp_{i,j,k}}{dt}=0$ for all $i,j,k$, by inspection.
		
		\textbf{Uniqueness}: Suppose $\frac{dp_{i,j,k}}{dt}=0$ for all $i,j,k$.  It follows immediately from \Cref{eq:GV_Model_DME} that
		\begin{align*}
			\frac{dp_{0,0,0}}{dt}={}&d\left[p_{1,0,0}+p_{0,1,0}+p_{0,0,1}\right]=0.\\
			\implies{}& p_{1,0,0}=p_{0,1,0}=p_{0,0,1}\equiv 0.\\
			\frac{dp_{1,0,0}}{dt}={}&d\left[2p_{2,0,0}+p_{1,1,0}+p_{1,0,1}\right]+\left(\frac{2}{\Omega}\right)p_{2,0,0}=0.\\
			\implies{}& p_{2,0,0}=p_{1,1,0}=p_{1,0,1}\equiv 0.\\
			\frac{dp_{1,1,0}}{dt}={}&d\left[2p_{2,1,0}+2p_{1,2,0}+p_{1,1,1}\right]+2\left(\frac{1+\alpha}{\Omega}\right)p_{2,1,0}+2\left(\frac{\beta+1}{\Omega}\right)p_{1,2,0}=0.\\
			\implies{}&p_{2,1,0}=p_{1,2,0}=p_{1,1,1}\equiv 0.\\
			\vdots{}&
		\end{align*}
		Continuing iteratively, it is clear that  $\pi(\mathbf{n})=0$ whenever $\mathbf{n}\neq 0$, while $p_{0,0,0}$ is not so constrained. 
		Normalization of the distribution enforces $\pi(\mathbf{n})=\delta(\mathbf{n})$.
	\end{proof}
	\section{Proof of Proposition 3.2}
	
	\label{app:Minimal_Model}
	Following \cite{Anderson2015a}, we summarize the elements of a chemical reaction network.
	The network comprises a set of $m$ species $\mathcal{S}$ (in our case, $\mathcal{S}=\{N_1,N_2,N_3\}$), a set of complexes $\mathcal{C}$, which are nonnegative integer linear combinations of species (for example, $N_1+N_2$ is the complex $y=(1,1,0)$; $2N_1$ is the complex $y'=(2,0,0)$, etc.), and a finite set $\mathcal{R}$ of reactions (e.g.~reaction 1 might be $N_1+N_2\to 2N_2$).
	A reaction network of this form has $\ell$ linkage classes, which are the number of connected components of the reaction network graph.
	We index the reactions $1,\dots,k,\dots,|\mathcal{R}|$. In a deterministic mass-action kinetics model, the reaction taking complex $y$ to complex $y'$ has rate $\kappa c^y,$ where $c\in\mathbb{R}^m_+$ is the vector of species concentrations, $\kappa$ is a molecular rate constant, and $c^y=\prod_{i=1}^mc_i^{y_i}$.    
	An equilibrium concentration $c_*$ for a deterministic network is ``complex-balanced" if for every complex $\eta\in\mathcal{C}$, the net production and consumption rates of $\eta$ are equal, i.e.
	\begin{equation}
		\sum_{\{k:\eta=y_k\}}\kappa_kc^{y_k}=\sum_{\{k:\eta=y'_k\}}\kappa_kc^{y'_k}
	\end{equation}
	where the LHS sums over source complexes and the RHS sums over product complexes.  
	Anderson and Kurtz \cite{Anderson2015a} further define a chemical reaction network $\{\mathcal{S},\mathcal{C},\mathcal{R}\}$ to be \emph{weakly reversible} if for any reaction $y_k\to y'_k \in\mathcal{R},$ there is a finite sequence of  reactions beginning with $y'_k$ as a source complex and ending with $y_k$ as a product complex, e.g.~$y'_k\to y_1\to y_2\to \ldots \to y_r\to y_k$.
	
	With this background, we are able to prove \cref{prop:minimal_model}:
	
	\textit{Let $\mathbf{N}=(N_1,N_2,N_3)$ be the population vector of the minimal model \Cref{eq:Minimal_Model_Reactions_1}-\cref{eq:Minimal_Model_Reactions_4}. 
		The reaction system \Cref{eq:Minimal_Model_Reactions_1}-\cref{eq:Minimal_Model_Reactions_4} has four distinct stationary distributions.
		Three may be expressed as component-wise stationary distributions of the form
		\begin{equation*}
			\pi(n_i)=\frac{\Omega^{n_i}}{n_i!(e^\Omega-1)}\delta(n_{i+1})\delta(n_{i+2}),
		\end{equation*}
		for $i\in\{1,2,3\}$, $n_i\ge 1$, with $\delta(x)$ being the distribution with unit probability at $x=0$, and with index addition taken cyclically on $\{1,2,3\}$.
		The fourth is $\pi(\mathbf{n})=\delta(\mathbf{n})$.}
	
	\begin{proof}
		The distribution $\pi(\mathbf{n})=\prod_{i=1}^3\delta(n_i)$, which represents complete extinction, is clearly an invariant distribution for the system \Cref{eq:Minimal_Model_Reactions_1}-\cref{eq:Minimal_Model_Reactions_4}. 
		It remains to show that the only other stationary distributions have the form \Cref{eq: Minimal Model Stationary Distribution}.
		
		We start by showing that neither the full three-dimensional system nor the two-dimensional subsystem admit complex balanced equilibria, while each one-dimensional subsystem does admit a complex balanced equilibrium (CBE).
		Without loss of generality, we will set $i=1$; the remaining stationary distribution follows by permutation of indices.
		First, we consider the full three-dimensional system.
		We write the reaction network from \Cref{eq:Minimal_Model_Reactions_1}-\cref{eq:Minimal_Model_Reactions_4} in the more compact form
		\begin{equation}
			\begin{gathered}
				N_1\rightleftharpoons2N_1 \qquad N_2\rightleftharpoons2N_2 \qquad N_3\rightleftharpoons2N_3 \\
				N_1\gets N_1+N_2\to N_2 \quad N_1\gets N_1+N_3\to N_3 \quad N_2\gets N_2+N_3\to N_3\\
			\end{gathered}
			\label{eq: Minimal_Model_3D_Reaction_Net}
		\end{equation}
		The system has nine complexes: $\mathcal{C}=\{N_1,N_2,N_3,2N_1,2N_2,2N_3,N_1+N_2,N_1+N_3,N_2+N_3\}$ and six linkage classes $\ell$ (the distinct connected components of the reaction network, displayed in \Cref{eq: Minimal_Model_3D_Reaction_Net}).
		The stoichiometric reaction vectors, representing the change in number of each species that results from each reaction, span the entire space of dimension $s=3$.
		The \emph{network deficiency} is  $\delta:=|\mathcal{C}|-\ell-s$;  
		for our system $\delta=9-6-3=0$. 
		The complex balanced equilibrium theorem \cite{anderson10,Anderson2015a} establishes that a zero-deficiency network has a CBE if and only if it is weakly reversible.
		The minimal reaction network \Cref{eq: Minimal_Model_3D_Reaction_Net} is not weakly reversible.  
		For example, the complex $y=N_1+N_2$ appears as the source complex in the reaction $N_1+N_2\to N_1$, but there is no reaction path leading from the product complex $y'=N_1$ back to $y$.
		We conclude that the network does not admit a CBE.  
		Consequently, there is no stationary distribution in which all three species have nonzero populations. 
		(Failure of weak reversibility coincides with the intuition that once the population enters a two-dimensional subspace on a coordinate plane, there is no reaction to bring the system back into the full three-dimensional space.)
		
		Next suppose, again WLOG, that $N_3$ is the first population to go extinct.
		The two dimensional subsystem $(N_1,N_2,0)$ is an absorbing set, within which the system has the reduced reaction network
		\begin{gather*}
			N_1\rightleftharpoons2N_1 \qquad N_2\rightleftharpoons2N_2 \\
			N_1\gets N_1+N_2\to N_2 \\
		\end{gather*}
		This reaction network has deficiency $\delta=5-3-2=0$.
		Invoking the complex balanced equilibrium theorem again, since this two-dimensional network also fails to be weakly reversible, it again does not admit a stationary distribution.
		
		Within the $(N_1,N_2,0)$ subsystem, either species could go extinct.  
		Suppose (again WLOG) that $N_2$ goes extinct next.
		Now the network reduces to the one-dimensional subsystem
		\begin{equation}
			N_1\rightleftharpoons2N_1
			\label{eq: Minimal_Model_1D_Reaction_Net}
		\end{equation}
		This reaction network also has zero deficiency, $\delta=2-1-1=0$, but unlike the previous cases, it is weakly reversible. 
		In this case, the stationary distribution theorem \cite{anderson10} establishes that the subsystem \Cref{eq: Minimal_Model_1D_Reaction_Net} has a unique stationary distribution which is a truncated Poisson distribution: 
		\begin{equation}
			\label{eq:Poisson} \pi(n_1)=\frac{\Omega^{n_1}}{n_1!\left(e^{\Omega}-1\right)}, \ n_1\in\mathbb{N}.
		\end{equation}
		
		The  structure of the three-population minimal  reaction network \Cref{eq:Minimal_Model_Reactions_1}-\cref{eq:Minimal_Model_Reactions_4} is invariant under permutation of the indices $\{1,2,3\}$.  
		Therefore, with system size $\Omega$, system \Cref{eq:Minimal_Model_Reactions_1}-\cref{eq:Minimal_Model_Reactions_4} admits precisely three non-degenerate stationary distributions, namely $\pi(n_i)\delta(n_{i+1})\delta(n_{i+2})$, where $\delta(n)$ is the  distribution with unit probability at $n=0$, and indicial addition is taken cyclically on $\{1,2,3\}$.  
		This completes the proof of \cref{prop:minimal_model}.
	\end{proof}
	
	\section{Formulation of Discrete First-Passage Problems}
	\label{app:Discrete_Operator}
	To construct the first-passage location and first-passage time problems for the stochastic minimal model as described in \cref{sec:ML_Transient_Behavior}, we first derive the infinitesimal generator matrix $\mathcal{L}$, with appropriate boundary conditions.
	To track when a particular species goes extinct, we require the coordinate planes $N_i=0$ to be absorbing, where $i\in\{1,2,3\}$.
	While in principle the populations in the minimal model are unbounded, we truncate the state space and obtain a finite-dimensional operator $\mathcal{L}$.
	To enforce conservation of probability, we impose reflecting boundary conditions on the planes $N_i=A$.
	The dimension of the resulting operator $\mathcal{L}$ is $(A+1)^3\times (A+1)^3$.
	We found that setting $A=2\Omega$ maintained a reasonable balance between accuracy and computational efficiency.
	
	Given the stochastic reaction net for the minimal model in \Cref{eq:Minimal_Model_Reactions_1}-\cref{eq:Minimal_Model_Reactions_4} and the boundary conditions specified above, we construct $\mathcal{L}$ following \cite{taylor98,Higham08,wilkinson18}.
	Each entry of $\mathcal{L}$ corresponds to a particular pair of states in the discrete system.
	Consider a state $i$ with population vector $\mathbf{N}=(N_1,N_2,N_3)$.
	If $i$ is not on an absorbing boundary, the corresponding row of $\mathcal{L}$ is given by
	\begin{gather}
		\left[\mathcal{L}\right]_{i,i-(1+2\Omega)^2}=\frac{\alpha}{\Omega}N_1N_3+\frac{\beta}{\Omega}N_2N_3+\frac{N_3(N_3-1)}{\Omega},\\
		\left[\mathcal{L}\right]_{i,i-(1+2\Omega)}=\frac{N_1(N_1-1)}{\Omega}+\frac{\alpha}{\Omega}N_1N_2+\frac{\beta}{\Omega}N_1N_3,\\
		\left[\mathcal{L}\right]_{i,i-1}=\frac{\beta}{\Omega}N_1N_2+\frac{N_2(N_2-1)}{\Omega}+\frac{\alpha}{\Omega}N_2N_3,\\
		\label{eq:Generator_Matrix_1}
		\left[\mathcal{L}\right]_{i,i+1}=rN_2\mathbbm{1}_{\{N_2<2\Omega\}}(N_2),\\
		\left[\mathcal{L}\right]_{i,i+(1+2\Omega)}=rN_1\mathbbm{1}_{\{N_1<2\Omega\}}(N_1),\\
		\label{eq:Generator_Matrix_3}
		\left[\mathcal{L}\right]_{i,i+(1+2\Omega)^2}=rN_3\mathbbm{1}_{\{N_3<2\Omega\}}(N_3),\\
		\left[\mathcal{L}\right]_{i,i}=-\sum_{j\neq i}\left[\mathcal{L}\right]_{i,j}.
	\end{gather}
	In \Cref{eq:Generator_Matrix_1}-\cref{eq:Generator_Matrix_3}, $\mathbbm{1}_{A}(X)$ is an indicator function that equals to unity if $X\in A$ and is zero otherwise.
	This indicator function enforces the reflecting boundary if $i$ is on one of the specified planes.
	If $i$ is on one of the absorbing boundaries, then the corresponding row of $\mathcal{L}$ is simply given by
	\begin{equation}
		\left[\mathcal{L}\right]_{i,i}=1.
	\end{equation}
	We index $\mathcal{L}$ in the method outlined above so that our implementation is compatible with \texttt{meshgrid} in MATLAB; see \url{https://github.com/nwbarendregt/StochasticHC}.
	By constructing $\mathcal{L}$ as specified above, we can formulate both the first-passage location problem in \Cref{eq:first-passage_location} and the first-passage time problem in \Cref{eq:first-passage_time} by modifying the right-hand side of the equation.
	
	\section*{Acknowledgments}
	Large-scale Monte Carlo simulations made use of the High Performance Computing Resource in the Core Facility for Advanced Research Computing at Case Western Reserve University.
	The second author acknowledges research support from Oberlin College.
	
	\section*{Code Availability}
	See \url{https://github.com/nwbarendregt/StochasticHC} for the MATLAB code used to generate all results and figures.
	
	\bibliographystyle{siamplain}
	\bibliography{References}

\begin{thebibliography}{10}

\bibitem{allen2010}
{\sc L.~J. Allen}, {\em An introduction to stochastic processes with
  applications to biology}, CRC press, 2010.

\bibitem{anderson10}
{\sc D.~F. Anderson, G.~Craciun, and T.~G. Kurtz}, {\em Product-form stationary
  distributions for deficiency zero chemical reaction networks}, Bulletin of
  mathematical biology, 72 (2010), pp.~1947--1970.

\bibitem{Anderson2015b}
{\sc D.~F. Anderson, B.~Ermentrout, and P.~J. Thomas}, {\em Stochastic
  representations of ion channel kinetics and exact stochastic simulation of
  neuronal dynamics}, Journal of computational neuroscience, 38 (2015),
  pp.~67--82.

\bibitem{Anderson2015a}
{\sc D.~F. Anderson and T.~G. Kurtz}, {\em Stochastic analysis of biochemical
  systems}, vol.~674, Springer, 2015.

\bibitem{benayoun2010}
{\sc M.~Benayoun, J.~D. Cowan, W.~van Drongelen, and E.~Wallace}, {\em
  Avalanches in a stochastic model of spiking neurons}, PLoS computational
  biology, 6 (2010), p.~e1000846.

\bibitem{bressloff2010}
{\sc P.~C. Bressloff}, {\em Stochastic neural field theory and the system-size
  expansion}, SIAM Journal on Applied Mathematics, 70 (2010), pp.~1488--1521.

\bibitem{browning2021}
{\sc A.~P. Browning, J.~A. Sharp, T.~Mapder, C.~M. Baker, K.~Burrage, and M.~J.
  Simpson}, {\em Persistence as an optimal hedging strategy}, Biophysical
  Journal, 120 (2021), pp.~133--142.

\bibitem{calvetti2012}
{\sc D.~Calvetti and E.~Somersalo}, {\em Computational mathematical modeling:
  an integrated approach across scales}, vol.~17, Siam, 2012.

\bibitem{cowan2016}
{\sc J.~D. Cowan, J.~Neuman, and W.~van Drongelen}, {\em Wilson--{C}owan
  equations for neocortical dynamics}, The Journal of Mathematical
  Neuroscience, 6 (2016), pp.~1--24.

\bibitem{cullins2015b}
{\sc M.~J. Cullins, J.~P. Gill, J.~M. McManus, H.~Lu, K.~M. Shaw, and H.~J.
  Chiel}, {\em Sensory feedback reduces individuality by increasing variability
  within subjects}, Current Biology, 25 (2015), pp.~2672--2676.

\bibitem{cullins2015}
{\sc M.~J. Cullins, K.~M. Shaw, J.~P. Gill, and H.~J. Chiel}, {\em Motor
  neuronal activity varies least among individuals when it matters most for
  behavior}, Journal of neurophysiology, 113 (2015), pp.~981--1000.

\bibitem{de2021}
{\sc A.~De~Candia, A.~Sarracino, I.~Apicella, and L.~de~Arcangelis}, {\em
  Critical behaviour of the stochastic {W}ilson-{C}owan model}, bioRxiv,
  (2021).

\bibitem{faugeras2015}
{\sc O.~Faugeras and J.~Inglis}, {\em Stochastic neural field equations: a
  rigorous footing}, Journal of mathematical biology, 71 (2015), pp.~259--300.

\bibitem{fox1994}
{\sc R.~F. Fox and Y.-n. Lu}, {\em Emergent collective behavior in large
  numbers of globally coupled independently stochastic ion channels}, Physical
  Review E, 49 (1994), p.~3421.

\bibitem{Gardiner09}
{\sc C.~Gardiner}, {\em Stochastic methods}, vol.~4, Springer Berlin, 2009.

\bibitem{gerdes2012}
{\sc K.~Gerdes and E.~Maisonneuve}, {\em Bacterial persistence and
  toxin-antitoxin loci}, Annual review of microbiology, 66 (2012),
  pp.~103--123.

\bibitem{gill2020thesis}
{\sc J.~P. Gill}, {\em Neural correlates of adaptive responses to changing load
  in feeding \textit{Aplysia}}, PhD thesis, Case Western Reserve University,
  2020.

\bibitem{gill2020}
{\sc J.~P. Gill and H.~J. Chiel}, {\em Rapid adaptation to changing mechanical
  load by ordered recruitment of identified motor neurons}, Eneuro, 7 (2020).

\bibitem{gillespie1977}
{\sc D.~T. Gillespie}, {\em Exact stochastic simulation of coupled chemical
  reactions}, The journal of physical chemistry, 81 (1977), pp.~2340--2361.

\bibitem{goldwyn2011}
{\sc J.~H. Goldwyn, N.~S. Imennov, M.~Famulare, and E.~Shea-Brown}, {\em
  Stochastic differential equation models for ion channel noise in
  {H}odgkin-{H}uxley neurons}, Physical Review E, 83 (2011), p.~041908.

\bibitem{goldwyn2011a}
{\sc J.~H. Goldwyn and E.~Shea-Brown}, {\em The what and where of adding
  channel noise to the {H}odgkin-{H}uxley equations}, PLoS computational
  biology, 7 (2011), p.~e1002247.

\bibitem{Higham08}
{\sc D.~J. Higham}, {\em Modeling and simulating chemical reactions}, SIAM
  review, 50 (2008), pp.~347--368.

\bibitem{Horn1972}
{\sc F.~Horn and R.~Jackson}, {\em General mass action kinetics}, Archive for
  rational mechanics and analysis, 47 (1972), pp.~81--116.

\bibitem{Kerr2002}
{\sc B.~Kerr, M.~A. Riley, M.~W. Feldman, and B.~J. Bohannan}, {\em Local
  dispersal promotes biodiversity in a real-life game of
  rock--paper--scissors}, Nature, 418 (2002), p.~171.

\bibitem{lotka1925}
{\sc A.~J. Lotka}, {\em Elements of physical biology}, Williams \& Wilkins,
  1925.

\bibitem{Lyttle2017}
{\sc D.~N. Lyttle, J.~P. Gill, K.~M. Shaw, P.~J. Thomas, and H.~J. Chiel}, {\em
  Robustness, flexibility, and sensitivity in a multifunctional motor control
  model}, Biological cybernetics, 111 (2017), pp.~25--47.

\bibitem{May1975}
{\sc R.~M. May and W.~J. Leonard}, {\em Nonlinear aspects of competition
  between three species}, SIAM journal on applied mathematics, 29 (1975),
  pp.~243--253.

\bibitem{orio2012}
{\sc P.~Orio and D.~Soudry}, {\em Simple, fast and accurate implementation of
  the diffusion approximation algorithm for stochastic ion channels with
  multiple states}, PLoS one, 7 (2012), p.~e36670.

\bibitem{park2018}
{\sc Y.~Park, K.~M. Shaw, H.~J. Chiel, and P.~J. Thomas}, {\em The
  infinitesimal phase response curves of oscillators in piecewise smooth
  dynamical systems}, European Journal of Applied Mathematics, 29 (2018),
  pp.~905--940.

\bibitem{pu2020}
{\sc S.~Pu and P.~J. Thomas}, {\em Fast and accurate langevin simulations of
  stochastic {H}odgkin-{H}uxley dynamics}, Neural Computation, 32 (2020),
  pp.~1775--1835.

\bibitem{pu2021}
{\sc S.~Pu and P.~J. Thomas}, {\em Resolving molecular contributions of ion
  channel noise to interspike interval variability through stochastic
  shielding}, Biological Cybernetics,  (2021), pp.~1--36.

\bibitem{purvis2000}
{\sc A.~Purvis, J.~L. Gittleman, G.~Cowlishaw, and G.~M. Mace}, {\em Predicting
  extinction risk in declining species}, Proceedings of the royal society of
  London. Series B: Biological Sciences, 267 (2000), pp.~1947--1952.

\bibitem{rabinovich2008}
{\sc M.~Rabinovich, R.~Huerta, and G.~Laurent}, {\em Transient dynamics for
  neural processing}, Science,  (2008), pp.~48--50.

\bibitem{Rabinovich2001}
{\sc M.~Rabinovich, A.~Volkovskii, P.~Lecanda, R.~Huerta, H.~Abarbanel, and
  G.~Laurent}, {\em Dynamical encoding by networks of competing neuron groups:
  winnerless competition}, Physical review letters, 87 (2001), p.~068102.

\bibitem{Reichenbach2006}
{\sc T.~Reichenbach, M.~Mobilia, and E.~Frey}, {\em Coexistence versus
  extinction in the stochastic cyclic {L}otka-{V}olterra model}, Physical
  Review E, 74 (2006), p.~051907.

\bibitem{safta2015}
{\sc C.~Safta, K.~Sargsyan, B.~Debusschere, and H.~N. Najm}, {\em Hybrid
  discrete/continuum algorithms for stochastic reaction networks}, Journal of
  Computational Physics, 281 (2015), pp.~177--198.

\bibitem{shaffer81}
{\sc M.~L. Shaffer}, {\em Minimum population sizes for species conservation},
  BioScience, 31 (1981), pp.~131--134.

\bibitem{Shaw2014PhDThesis}
{\sc K.~M. Shaw}, {\em Dynamical Architectures for Controlling Feeding in
  \textit{Aplysia californica}}, PhD thesis, Case Western Reserve University,
  2014.

\bibitem{shaw15}
{\sc K.~M. Shaw, D.~N. Lyttle, J.~P. Gill, M.~J. Cullins, J.~M. McManus, H.~Lu,
  P.~J. Thomas, and H.~J. Chiel}, {\em The significance of dynamical
  architecture for adaptive responses to mechanical loads during rhythmic
  behavior}, Journal of computational neuroscience, 38 (2015), pp.~25--51.

\bibitem{shaw2012}
{\sc K.~M. Shaw, Y.-M. Park, H.~J. Chiel, and P.~J. Thomas}, {\em Phase
  resetting in an asymptotically phaseless system: On the phase response of
  limit cycles verging on a heteroclinic orbit}, SIAM Journal on Applied
  Dynamical Systems, 11 (2012), pp.~350--391.

\bibitem{Sinervo1996}
{\sc B.~Sinervo and C.~M. Lively}, {\em The rock--paper--scissors game and the
  evolution of alternative male strategies}, Nature, 380 (1996), p.~240.

\bibitem{Strang2019}
{\sc A.~G. Strang, K.~C. Abbott, and P.~J. Thomas}, {\em How to avoid an
  extinction time paradox}, Theoretical Ecology,  (2019).

\bibitem{taylor98}
{\sc H.~M. Taylor and S.~Karlin}, {\em An introduction to stochastic modeling},
  Academic Press, New York, 1998.

\bibitem{van92}
{\sc N.~G. Van~Kampen}, {\em Stochastic processes in physics and chemistry},
  vol.~1, Elsevier, 1992.

\bibitem{varona2002}
{\sc P.~Varona, M.~I. Rabinovich, A.~I. Selverston, and Y.~I. Arshavsky}, {\em
  Winnerless competition between sensory neurons generates chaos: A possible
  mechanism for molluscan hunting behavior}, Chaos: An Interdisciplinary
  Journal of Nonlinear Science, 12 (2002), pp.~672--677.

\bibitem{Vellela2007}
{\sc M.~Vellela and H.~Qian}, {\em A quasistationary analysis of a stochastic
  chemical reaction: Keizer’s paradox}, Bulletin of mathematical biology, 69
  (2007), pp.~1727--1746.

\bibitem{volterra1926}
{\sc V.~Volterra}, {\em Variazioni e fluttuazioni del numero d'individui in
  specie animali conviventi}, 1926.

\bibitem{webster2020}
{\sc V.~A. Webster-Wood, J.~P. Gill, P.~J. Thomas, and H.~J. Chiel}, {\em
  Control for multifunctionality: bioinspired control based on feeding in
  \textit{{A}plysia californica}}, Biological Cybernetics, 114 (2020),
  pp.~557--588.

\bibitem{wilkinson18}
{\sc D.~J. Wilkinson}, {\em Stochastic modelling for systems biology}, CRC
  press, 2018.

\bibitem{Wilson1972}
{\sc H.~R. Wilson and J.~D. Cowan}, {\em Excitatory and inhibitory interactions
  in localized populations of model neurons}, Biophysical journal, 12 (1972),
  pp.~1--24.

\bibitem{Wilson1973}
{\sc H.~R. Wilson and J.~D. Cowan}, {\em A mathematical theory of the
  functional dynamics of cortical and thalamic nervous tissue}, Kybernetik, 13
  (1973), pp.~55--80.

\bibitem{xue2017}
{\sc C.~Xue and N.~Goldenfeld}, {\em Coevolution maintains diversity in the
  stochastic “kill the winner” model}, Physical review letters, 119 (2017),
  p.~268101.

\bibitem{yahalom2019}
{\sc Y.~Yahalom, B.~Steinmetz, and N.~M. Shnerb}, {\em Comprehensive phase
  diagram for logistic populations in fluctuating environment}, Physical Review
  E, 99 (2019), p.~062417.

\end{thebibliography}
\end{document}